\documentclass{PoS}

\usepackage{amsmath,amssymb}
\usepackage{graphicx}

\newcommand{\Comments}[1]{}

\newcommand{\id}{{\bf 1}}
\newcommand{\arxiv}[1]{{arXiv:#1}}
\newcommand{\be}{\begin{equation}}
\newcommand{\ee}{\end{equation}}
\newcommand{\ba}{\begin{align}}
\newcommand{\ea}{\end{align}}
\newcommand{\nn}{\nonumber}

\newcommand{\chibar}{{\bar{\chi}}}

\newcommand{\Fint}{{\cal D}}

\newcommand{\Mhat}{\hat{M}}

\title{Strong coupling analysis of Aoki phase in Staggered-Wilson fermions\thanks{YITP-12-83, KUNS-2419, RIKEN-MP-56}}

\ShortTitle{Strong coupling analysis of Aoki phase in Staggered-Wilson fermions}

\author{\speaker{Takashi Z. Nakano} \\ 
        Department of Physics, Yukawa Institute for Theoretical Physics, Kyoto University,
        Kyoto 606-8502, Japan \\
        E-mail: \email{tnakano@yukawa.kyoto-u.ac.jp}}

\author{Tatsuhiro Misumi\\
        Physics Department, Brookhaven National Laboratory,
        Upton, New York 11973, USA
        }

\author{Taro Kimura\\
        Mathematical Physics Laboratory, RIKEN Nishina Center, 
        Saitama 351-0198, Japan
        }

\author{Akira Ohnishi\\
        Yukawa Institute for Theoretical Physics, Kyoto University,
Kyoto 606-8502, Japan \\
        }

\abstract{We study strong-coupling lattice QCD with staggered-Wilson fermions,
with an emphasis on the possibility of spontaneous parity breaking.
We perform effective potential analysis in the strong-coupling limit. 
From gap equations we find the pion condensate becomes nonzero
in some range of a mass parameter,
which indicates the existence of the parity-broken phase. 
We also find massless pions and PCAC relations
around the second-order phase boundary.
These results suggest that we can take the chiral limit by 
tuning a mass parameter in lattice QCD with staggered-Wilson fermions
as with the Wilson fermion. 
}

\FullConference{The 30th International Symposium on Lattice Field Theory\\
		June 24-29, 2012\\
		Cairns, Australia}

\begin{document}

\section{Introduction}
\label{sec:Intro}

Since the dawn of lattice field theory, the doubling problem has been a notorious
obstacle for lattice simulations \cite{NN}. 
Although several prescriptions for this problem have been proposed, 
including Wilson \cite{Wilson}, staggered \cite{KS}, domain-wall 
\cite{DW} and overlap fermions \cite{OV},
they have their individual flaws such as high numerical cost or the undesirable 
number of flavors. Recently, a new possibility of lattice fermion constructions 
was pointed out, which is called staggered-Wilson or staggered-Overlap \cite{StW}.
It is constructed by introducing staggered versions of generalized 
Wilson terms \cite{CKM} into staggered fermions.
One possible advantage of this formulation is that it would improve 
the taste breaking \cite{Steve} and reduce numerical costs of overlap fermions \cite{PdF}.
As with the case of Wilson fermion, it is quite important to study 
the parity phase structure (Aoki phase) \cite{AP} in staggered-Wilson fermions. 
The phase structure for the lattice Gross-Neveu model with 
staggered-Wilson fermions were studied in Ref.~\cite{GNStW, MCKNO} 
and by using hopping parameter expansion in strong-coupling lattice QCD \cite{MCKNO, MKNO}. 
In this proceedings, we investigate strong-coupling lattice QCD \cite{KaS} 
with emphasis on parity-phase structure \cite{MKNO} for staggered-Wilson fermions.
We perform effective potential analysis for meson fields in the strong-coupling limit.
The gap equations show that the pion condensate becomes nonzero in some range of a mass parameter. We also study meson masses around the second-order phase boundary, 
and find massless pions and PCAC relation.


\section{Staggered-Wilson fermions}
\label{sec:SWF}

We first introduce ``flavored mass terms" or ``taste-dependent mass terms" 
which are generalization of the Wilson term \cite{CKM}.
It is shown that there are two types of such terms \cite{StW},
but we focus only on one of them, which we call Adams type, 
since it has sufficient spacetime symmetries \cite{Steve}.
It is composed of four hopping terms as
\begin{equation}  
M_A= \epsilon\sum_{sym} \eta_{1}\eta_{2}\eta_{3}\eta_{4}
C_{1}C_{2}C_{3}C_{4}
= (\id \otimes \gamma_{5}) + \mathcal{O}(a)
\ ,
\label{AdamsM}
\end{equation}
with
$(\epsilon)_{xy}=(-1)^{x_{1}+...+x_{4}}\delta_{x,y}
,(\eta_{\mu})_{xy}=(-1)^{x_{1}+...+x_{\mu-1}}\delta_{x,y}
,C_{\mu}=(V_{\mu}+V_{\mu}^{\dag})/2
,(V_{\mu})_{xy}=U_{\mu,x}\delta_{y,x+\hat{\mu}}
$.
Added to usual staggered fermion actions, this term lifts the degeneracy 
of four tastes and ends up with two positive-mass flavors and two negative-mass flavors.
The two branches correspond to $+1$ and $-1$ eigenvalues of 
$\gamma_{5}$ in the taste space.
We note that $M_{A}$ is also derived from the flavored
mass terms for naive fermions through spin-diagonalization as shown in \cite{CKM}.   
Then the Adams-type staggered-Wilson fermion action is given by
\begin{align}  
S_{\rm A}\,&=\,\sum_{xy}\bar{\chi}_{x}[\eta_{\mu}D_{\mu}
+r(1+M_A)+M]_{xy}\chi_{y}
\ ,
\label{AdamsS}
\end{align}
where $D_{\mu}=(V_{\mu}-V^\dagger_{\mu})/2$.
Here $\chi$, $r$, and $M$ are the quark field, the Wilson parameter,
and the usual taste-singlet mass ($M=M\delta_{x,y}$), respectively. 
In lattice QCD simulations with these fermions, the mass parameter $M$ will be
tuned to take a chiral limit as in the Wilson fermion. 
By substituting this Dirac kernel with $-1<M<0$,
we obtain a two-flavor overlap fermion.

\section{Effective Potential Analysis}
\label{sec:EPA}
In this section, we consider the effective potential of meson
fields for $SU(N)$ lattice gauge theory with staggered-Wilson fermions.
In the strong-coupling limit and the large $N$ limit, 
the effective action can be exactly derived by integrating the link variables \cite{AP, KaS}.
Then, by solving a saddle point equation, we can investigate a vacuum
and find meson condensations.

\begin{figure}
 \begin{center}
  \includegraphics[height=6cm]{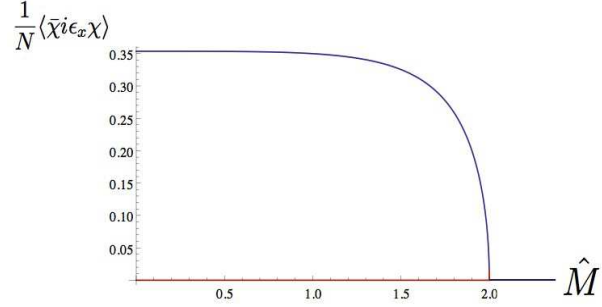}\,\,\,\,
   \end{center}
 \caption{The pion condensate undergoes second-order phase transition.}
 \label{trans1}
\end{figure}

In the strong-coupling limit we can drop the plaquette action.  
Then the partition function for meson fields
$\mathcal{M}_x=(\chibar_x \chi_x)/N$ with the source $J_{x}$ is given by
\begin{align}
Z(J) &= \displaystyle \int \Fint \left[ \chi, \chibar, U \right]
 \exp \left[ N \sum_x J_x \mathcal{M}_x + S_F \right]
\ .
\label{ZJ}
\end{align}
where $S_{F}$ stands for the fermion action. 
In this case, $S_F$ is the Adams type 
staggered-Wilson action Eq.~(\ref{AdamsS}). $N$ stands for the number of color. 
We here consider the effective action up to $\mathcal{O}(\mathcal{M}^{3})$
for the meson field $\mathcal{M}$.
A method to perform the link variable integral for multi-hopping terms is 
developed in \cite{MKNO}.
By using this method, 
the effective partition function for the meson field is given by
\begin{align}
Z(J)
&= \displaystyle \int \Fint \mathcal{M} \exp \left[ N \left( \sum_x J_x \mathcal{M}_x 
+ S_{\mathrm{eff}}(\mathcal{M}) \right) \right] 
\ , \\
S_{\mathrm{eff}}(\mathcal{M})
&= \sum_x \left( \Mhat \mathcal{M}_x - \ln \mathcal{M}_x \right) + \sum_x W(\Lambda)
\ ,
\label{EAc1}
\end{align}
where we denote $\Mhat$ as the shifted mass parameter $\Mhat = M+r$.
$W(\Lambda)$ is,
\begin{align}
\sum_x W(\Lambda)
&= - \sum_x \left[ \left( 1-4 \Lambda_x \right)^{1/2} 
- 1 - \ln  \left[ \displaystyle \frac {1+
\left( 1-4 \Lambda_x \right)^{1/2}}{2} \right] \right]
\ ,
\end{align}
where $\Lambda_x$ is, 
\begin{align}
\Lambda_x
&=
\displaystyle \frac {1}{16}  
\biggl[ 
 \sum_\mu \mathcal{M}_x \mathcal{M}_{x+\hat{\mu}} 
 + \displaystyle \frac {1}{3} \sum_{\mu \neq \nu} \mathcal{M}_{x+\hat{\mu}} \mathcal{M}_{x+\hat{\mu}+\hat{\nu}} 
\ \nn \\
& + \displaystyle \frac {1}{6} \sum_{\mu \neq \nu \neq \rho} \mathcal{M}_{x+\hat{\mu}+\hat{\nu}} \mathcal{M}_{x+\hat{\mu}+\hat{\nu}+\hat{\rho}} 
  + \displaystyle \frac {1}{6} \sum_{\mu \neq \nu \neq \rho \neq \sigma} \mathcal{M}_{x+\hat{\mu}+\hat{\nu}+\hat{\rho}} \mathcal{M}_{x+\hat{\mu}+\hat{\nu}+\hat{\rho}+\hat{\sigma}} 
\biggr]
\ \nn \\
&-\left( \displaystyle \frac {r}{4! \cdot 16} \right)^2 
\sum_{\mu \neq \nu \neq \rho \neq \sigma} 
\left( 
2 \mathcal{M}_x \mathcal{M}_{x+\hat{\mu}+\hat{\nu}+\hat{\rho}+\hat{\sigma}} 
+4 \mathcal{M}_{x+\hat{\mu}} \mathcal{M}_{x+\hat{\nu}+\hat{\rho}+\hat{\sigma}} 
+2 \mathcal{M}_{x+\hat{\mu}+\hat{\nu}} \mathcal{M}_{x+\hat{\rho}+\hat{\sigma}} 
\right)
\ .
\end{align}
Then, we can derive the partition function in large $N_c$ as, 
\begin{align}
Z(J=0)
&= \displaystyle \int \Fint \mathcal{M} \exp \left[ N S_{\mathrm{eff}}(\mathcal{M}) \right]
\ \nn \\
& \sim \exp \left[ N S_{\mathrm{eff}}(\bar{\mathcal{M}}) \right], 
\quad \left( N \rightarrow \infty \right)
\ ,
\end{align}
We assume only scalar (chiral) $\sigma$ and pseudoscalar (pion) 
$\pi$ fields as
$\bar{\mathcal{M}}_{x}
= \sigma + i \epsilon_x \pi 
= \Sigma e^{i\epsilon_x \theta}
$.
By substituting this form of the meson field into Eq.~(\ref{EAc1}),
we derive the effective action for the $\Sigma$ and $\theta$,
\begin{align}
S_{\mathrm{eff}}(\bar{\mathcal{M}})
&= \Mhat \sum_x \Sigma \cos \theta - \sum_x \ln \Sigma
\ \nn \\
&- \sum_x \left[ \left( 1- 4 \cdot 2 \Sigma^2 \sin^2 \theta \right)^{1/2} 
 - \ln \left[ \displaystyle \frac{1+\left( 1- 4 \cdot 2 \Sigma^2 \sin^2 
 \theta \right)^{1/2}}{2} \right] \right]
\ .
\end{align}
We ignore the irrelevant constant.
From the translational invariance, we factorize the four-dimensional 
volume $V_4$ from the effective action as 
$S_{\mathrm{eff}}(\bar{M})=-V_4 V_\mathrm{eff}(\Sigma,\theta)$.
Then the effective potential $V_{\mathrm{eff}}$ is given by
\begin{align}
V_{\mathrm{eff}}(\Sigma,\theta)
&= - \Mhat \Sigma \cos \theta + \ln \Sigma
\ \nn \\
&+ \left[ \left( 1- 4 \cdot 2 \Sigma^2 \sin^2 \theta \right)^{1/2} 
 - \ln \left[ \displaystyle \frac{1+\left( 1- 4 \cdot 2 \Sigma^2 
 \sin^2 \theta \right)^{1/2}}{2} \right] \right]
\ .
\label{effS1}
\end{align}
where $\hat{M}=M+r$. We take $r=16\sqrt{3}$ for simplicity
and ignore the irrelevant constant in the potential.
Now let us look into the vacuum structure of this effective potential
by solving the saddle point conditions 
$\partial V_{\mathrm{eff}}(\Sigma,\theta)/\partial \Sigma=0,\partial V_{\mathrm{eff}}(\Sigma,\theta)/\partial \theta=0$.
For $\Mhat^2 > 4$ or, equivalently, $M > -16\sqrt{3}+2,\,\, M<-16\sqrt{3}-2$,
there is only the chiral condensate as
\begin{align}
\displaystyle \frac {1}{N} \langle \chibar \chi \rangle 
= \Sigma \cos \theta = \displaystyle \frac {1}{\Mhat}
, \quad
\displaystyle \frac {1}{N} \langle \chibar i \epsilon_x \chi \rangle 
= \Sigma \sin\theta = 0
\ .
\end{align}
For $\Mhat^2 < 4$ or, equivalently, $-16\sqrt{3}-2<M<-16\sqrt{3}+2$,
a finite pion condensate appears and breaks the parity symmetry spontaneously.
\begin{align}
\displaystyle \frac {1}{N} \langle \chibar \chi \rangle &=
\bar{\Sigma} \cos \bar{\theta} = \displaystyle \frac {\Mhat}{8-\Mhat^2}
, \quad 
\displaystyle \frac {1}{N} \langle \chibar i \epsilon_x \chi \rangle 
= \bar{\Sigma} \sin \bar{\theta} =  
\pm \displaystyle \frac {\sqrt{ 2(4-\Mhat^2)}}{8-\Mhat^2}
\ .
\label{picond}
\end{align}
The sign of the pion condensate Eq.~(\ref{picond}) reflects the $Z_2$ parity symmetry of the theory.
The critical mass parameter $M_{c}=-16\sqrt{3}\pm 2$ and the range for the Aoki phase
$-16\sqrt{3}-2<M<-16\sqrt{3}+2$ are consistent with those of the hopping 
parameter expansion \cite{MCKNO, MKNO}. 
These results strongly suggest the existence of the parity-broken phase in the lattice
QCD at least in the strong-coupling limit.
Figure~\ref{trans1} shows the pion condensate,
indicating that the phase transition is second-order.

We can also derive the mass spectrum of the mesons by expanding the 
effective action $S_{\mathrm{eff}}(\mathcal{M})$ to the quadratic terms of  
the meson excitation field $\Pi_x=\mathcal{M}_x-\bar{\mathcal{M}}_x$.
We here concentrate on the pion mass in the parity-symmetric phase,
since 
we can take the chiral limit by tuning the mass parameter
from the parity-symmetric phase to the critical line, 
In the parity-symmetric phase ($\Mhat^{2}>4$), 
the quadratic part of the effective action is given by
\begin{align}
S_\mathrm{eff}(\mathcal{M}) - S_\mathrm{eff}(\bar{\mathcal{M}})
&= \sum_{x,y} \Pi_x S_\mathrm{eff}^{(2)} (x,y) \Pi_y
\nonumber\\
&= \int_{-\pi}^\pi \displaystyle \frac{d^4p}{(2\pi)^4} 
 \Pi(-p) \mathcal{D} \Pi(p)
\ ,
\end{align}
where $\Pi(p)$ is the Fourier component of $\Pi_x$,
and
\begin{align}
\mathcal{D}
&= \displaystyle \frac {1}{2\Sigma^2} 
 + \left[ \displaystyle \frac{1}{4} \sum_\mu \cos p_\mu 
  - \displaystyle \frac{1}{24} 
    \sum_{\mu \neq \nu \neq \rho \neq \sigma} \cos p_{\mu + \nu + \rho + \sigma}  
\right] 
\ ,
\label{ApoD}
\end{align}
with $p_{\mu + \nu + \rho + \sigma}\equiv p_{\mu} + p_{\nu} + p_{\rho} + p_{\sigma}$.
We obtain the pion mass by solving $\mathcal{D}=0$ at $p=(i m_\pi a + \pi, \pi,\pi,\pi)$ as
\begin{equation}
\cosh(m_{\pi}a) = 1+{2\Mhat^{2}-8\over{5}}
\ .
\label{ApiM}
\end{equation}
We find that these results are  
consistent with the hopping parameter expansion \cite{MCKNO, MKNO}:
the pion becomes massless at
the critical mass $\Mhat^{2}=4$, which indicates that a second-order 
phase transition occurs between the parity-symmetric and broken phases in the 
strong-coupling limit. By defining the quark mass as $m_{q}a=\Mhat-\Mhat_{c}$,
we find an approximate PCAC relation near the critical mass,
\begin{equation}
(m_{\pi}a)^{2} = {16\over{5}}m_{q}a+\mathcal{O}(a^{2})
\ .
\label{ApiM2}
\end{equation}
We can also study a case for nonzero spatial
momenta by considering $p=(iEa+\pi,p_{1}a+\pi,p_{2}a+\pi,p_{3}a+\pi)$ 
in Eq.~(\ref{ApoD}). By using the pion mass Eq.~(\ref{ApiM2}) and renormalizing 
the Dirac operator as $-{8\over{5}}\mathcal{D}\to\mathcal{D}$, we show that
Eq.~(\ref{ApoD}) results in the Lorentz-covariant form up to $\mathcal{O}(a)$ discretization errors,
\begin{equation}
\mathcal{D}=(E^{2}-{\bf p}^{2}-m_{\pi}^{2})a^{2}+\mathcal{O}(a^{3})
\ ,
\label{ALorentz}
\end{equation}
with ${\bf p}^{2}=p_{1}^{2}+p_{2}^{2}+p_{3}^{2}$.

We here discuss the possibility of other condensations.
For this purpose, we consider a general form 
of the meson field,
\begin{equation} 
\bar{\mathcal{M}}_{x}=\sigma+i\epsilon_{x}\pi+\sum_{\mu}(-1)^{x_{\mu}}v_{\mu}
+\sum_{\mu}i\epsilon_{x}(-1)^{x_{\mu}}a_{\mu}
+\sum_{\mu>\nu}(-1)^{x_{\mu}+x_{\nu}}t_{\mu\nu}
\ ,
\label{GeM}
\end{equation}
where we define the vector, axial-vector, and tensor meson fields
as $v_{\mu}$, $a_{\mu}$, and $t_{\mu}$, respectively.
We can easily show that there is no other condensate by substituting this general 
form Eq.~(\ref{GeM}) into the meson action $S_{\mathrm{eff}}(\bar{\mathcal{M}})$. 
Thus we conclude that the vacuum we obtained is a true one.

\section{Discussion of the two-flavor properties in continuum limit}
\label{sec:Tf}
We now discuss the properties of flavors and pions for Adams-type
staggered-Wilson fermions near the continuum limit.
It has two flavors for each branch in the first place,
and there is a possibility of
simulating the two-flavor QCD just by using a single lattice fermion.
In lattice QCD with a single Adams-type fermion,
the pion condensate is likely to be given by
$\langle \chibar i \epsilon_x \chi \rangle  = 0$
in the symmetric phase and 
$\langle \chibar i \epsilon_x \chi \rangle \neq 0$ 
in the symmetry-broken phase where no explicit flavor indices appear.
We note that the two flavors contained in the Adams-type fermion 
have exact $U(1)$ baryon symmetry, but no exact 
$SU(2)$ flavor (taste) symmetry even near the chiral limit
because of the taste-mixing in the original staggered fermions.
This fact leads to the following properties:
In the parity-broken phase, there is no massless NG boson 
since there is no continuous symmetry to be broken
unlike the usual Wilson fermion. 
On the other hand, in the parity-symmetric phase and on the phase boundary, 
we will find correct chiral properties (PCAC) only for the $U(1)$ subgroup 
but not for the whole $SU(2)$ at finite lattice spacing.
Nevertheless, since the Lorentz symmetry and other requisite 
symmetries are expected to recover correctly in the parity-symmetric 
phase for the Adams type,
the universality class indicates that the $SU(2)$ flavor (taste) symmetry 
is recovered in the continuum limit.
Thus we believe that we can correctly describe the two-flavor QCD 
as long as we take a continuum limit.

We note that we may have a further bonus in this formulation.
$SU(2)$ flavor symmetry breaking does not necessarily imply the 
mass splitting of three pions. The mass degeneracy of the three pions 
depends on the discrete flavor symmetry in the pion sector:
if this symmetry is large enough to have a degenerate pion triplet, 
we have three degenerate pions even at the finite lattice spacing which
become massless in the chiral limit (on the boundary).
Recently Ref.~\cite{Steve} has reported that classification of pion operators 
from the transfer matrix symmetry indicates three degenerate pions even at 
finite lattice spacing in the Adams fermion. 
It is also notable that, if we start with the improved staggered action as HISQ \cite{HISQ}, 
the $SU(2)$ flavor breaking in the Adams fermion is expected to be improved and 
to yield better chiral properties.


\section{Summary and Discussion}
\label{sec:SD}

In this proceedings, we have investigated 
strong-coupling lattice QCD with the Adams-type 
staggered-Wilson fermions, with emphasis on the parity-broken phase (Aoki phase) structure. 
We have performed an effective potential analysis for meson fields
in the strong-coupling limit.
In some range of the fermion mass parameter,
we find a nontrivial parity-broken phase emerges.
We also show that the pions become massless on the phase boundary and 
pion mass obeys PCAC relations.
These results suggest that we can take a chiral limit by 
tuning a mass parameter in lattice QCD with staggered-Wilson 
fermions as with the Wilson fermion.

In the future work, we can also study the contribution
from some of higher meson fields,
the detailed mass 
spectrum of the mesons and the possibility of other small condensations 
in the Aoki phase.


\begin{acknowledgments}
TN and TK are supported by Grants-in-Aid for the Japan Society 
for Promotion of Science (JSPS) Research Fellows
(Nos.
22-3314, 
23-593.
).
TM is supported by Grant-in-Aid for the Japan Society for Promotion of Science (JSPS) Postdoctoral
Fellows for Research Abroad (24-8).
This work is supported in part by the Grants-in-Aid for Scientific Research
from JSPS and
the Ministry of Education, Culture, Sports, Science and Technology (MEXT)
(Nos. 
09J01226, 
10J03314, 
11J00593, 
23340067, 
24340054, 
24540271, 
and Innovative Areas (No. 2404: 24105001, 24105008) 
), 
by the Yukawa International Program for Quark-hadron Sciences,
and
by the Grant-in-Aid for the global COE program ``The Next Generation
of Physics, Spun from Universality and Emergence" from MEXT.
\end{acknowledgments}


\begin{thebibliography}{99}

\bibitem{NN}
H.~B.~Nielsen and M.~Ninomiya, Nucl. Phys. {\bf B185}, 20 (1981); 
Nucl. Phys. {\bf B193} 173 (1981);
Phys. Lett. B {\bf 105} 219 (1981).

\bibitem{Wilson}
K.~G.~Wilson, 
Quarks and strings on a lattice, in "Gauge Theories and Modern Field Theory"
MIT Press, Cambridge, 1975;
Quarks and strings on a lattice, in "New phenomena in subnuclear physics"
Plenum Press, New York, 1977. 

\bibitem{KS}
J.~B.~Kogut and L.~Susskind, Phys. Rev. D {\bf 11}, 395 (1975);
L.~Susskind, Phys. Rev. D {\bf 16}, 3031 (1977);


\bibitem{DW}
D.~B.~Kaplan, Phys. Lett. B {\bf 288}, 342 (1992);


\bibitem{OV}
N.~Neuberger, Phys. Lett. B {\bf 427}, 353 (1998).  

\bibitem{StW}
D.~H.~Adams, Phys. Rev. Lett. {\bf 104}, 141602 (2010);
Phys. Lett. B {\bf 699}, 394 (2011);
C.~Hoelbling, Phys. Lett. B {\bf 696}, 422 (2011).

\bibitem{CKM}
M.~Creutz, T.~Kimura, and T.~Misumi, J. High Energy Phys. {\bf 1012} (2010) 041; 
S.~Durr and G.~Koutsou, Phys. Rev. D {\bf 83}, 114512 (2011).

\bibitem{PdF}
P. de Forcrand, A. Kurkela and M. Panero, PoS Lattice2010 (2011) 080  
; J. High Energy Phys. {\bf 1204} (2012) 142.

\bibitem{AP}
S.~Aoki, Phys. Rev. D {\bf 30}, 2653 (1984); 
M.~Creutz, [\arxiv{hep-lat/9608024}];
S.~Sharpe and R.~Singleton.~Jr, Phys. Rev. D {\bf 58}, 074501 (1998).

\bibitem{GNStW}
M.~Creutz, T.~Kimura, and T.~Misumi,  Phys. Rev. D {\bf 83}, 094506 (2011).

\bibitem{MCKNO}
T.~Misumi, M.~Creutz, T.~Kimura, T.~Z~Nakano and A.~Ohnishi,
PoS Lattice2011 (2011) 108.

\bibitem{MKNO}
T.~Misumi, T.~Z~Nakano, T.~Kimura and A.~Ohnishi,
Phys. Rev. D {\bf 86}, 034501 (2012) .

\bibitem{KaS}
N.~Kawamoto and J.~Smit, Nucl. Phys. {\bf B192}, 100 (1981). 

\bibitem{Steve}
S.~Sharpe, talk at YIPQS-HPCI workshop ``New-Type of Fermions on the Lattice" (2012),
http://www2.yukawa.kyoto-u.ac.jp/ws/2011/newtype/Talk-slides/sharpe-kyoto12-1.pdf
\\ (to be published).

\bibitem{HISQ}
E.~Follana, Q.~Mason, C.~Davies, K.~Hornbostel, G.~P.~Lepage, 
J.~Shigemitsu, H.~Trottier, and K.~Wong,
Phys. Rev. D {\bf 75}, 054502 (2007).



\end{thebibliography}
\end{document}